\documentclass[twocolumn,prd,nofootinbib,superscriptaddress,longbibliography]{revtex4} %
\newcommand\ForInternalReference[1]{}
\newcommand\SkipForEarlyCirculation[1]{}

\newcommand\SkipPP[1]{}
\usepackage[colorlinks=true,citecolor=blue,urlcolor=blue]{hyperref}
\usepackage{verbatim}
\usepackage{graphicx}
\usepackage{dcolumn}
\usepackage{bm}
\usepackage{color}
\usepackage{xcolor}
\usepackage{url}
\usepackage{float}
\usepackage{multirow}
\usepackage{adjustbox}
\usepackage[normalem]{ulem}
\usepackage{bm}
\usepackage{placeins}
\usepackage{afterpage}
\usepackage{amsmath,amssymb}
\usepackage{acronym}
\usepackage{booktabs}      
\usepackage{tabularx}      
\usepackage{array}         

\newcommand\optional[1]{}
\acrodef{NR}[NR]{Numerical Relativity}

\usepackage{color}
\definecolor{amber}{rgb}{1.0, 0.75, 0.0}
\definecolor{orange}{rgb}{1.0, 0.5, 0.0}
\definecolor{amaranth}{rgb}{0.9, 0.17, 0.31}

\graphicspath{{./Figures/}}

\def\ltsima{$\; \buildrel < \over \sim \;$}
\def\simlt{\lower.5ex\hbox{\ltsima}}
\def\gtsima{$\; \buildrel > \over \sim \;$}
\def\simgt{\lower.5ex\hbox{\gtsima}}

\def\eos#1{equation of state#1 (EOS#1)\gdef\eos{EOS}}
\def\QNM#1{quasi-normal mode#1 (QNM#1)\gdef\QNM{QNM}}
\def\ns#1{neutron star#1 (NS#1)\gdef\ns{NS}}
\def\gw#1{gravitational wave#1 (GW#1)\gdef\gw{GW}}
\def\bh#1{black hole#1 (BH#1)\gdef\bh{BH}}
\def\bbh#1{binary black hole#1  (BBH#1)\gdef\bbh{BBH}}
\def\bns#1{binary neutron star#1 (BNS#1)\gdef\bns{BHS}}
\def\bhns#1{black hole - neutron star#1 (BHNS#1)\gdef\bhns{BHNS}}
\def\nsbh#1{neutron star - black hole#1 (NSBH#1)\gdef\nsbh{NSBH}}
\def\nr#1{numerical relativity#1 (NR#1)\gdef\nr{NR}}

\newcommand{\UT}{\affiliation{Center for Gravitational Physics, The University of Texas at Austin, Austin, Texas 78712, USA}}

\begin{document}

\title{Measuring eccentricity and addressing waveform systematics in GW231123} 
\author{Aasim Jan}
\UT
\author{Sophia Nicolella}
\UT
\author{Deirdre Shoemaker}
\UT
\author{Richard O'Shaughnessy}
\affiliation{Center for Computational Relativity and Gravitation, Rochester Institute of Technology, Rochester, New York 14623, USA}
\begin{abstract}
The gravitational-wave event GW231123\_135430  is the heaviest binary black hole system observed by the LIGO--Virgo--KAGRA Collaboration to date, with the initial analysis indicating the individual black hole masses lie within or above the theorized pair-instability mass gap of roughly $60$--$130\,M_\odot$. The inference further suggests that both black holes possess high spins, measured to be $0.90^{+0.10}_{-0.19}$ and $0.80^{+0.20}_{-0.51}$. Therefore, the observation of this event suggests the formation of black holes from channels beyond the standard stellar collapse. However, different waveform models yield significantly different parameter estimates, possibly due to missing physics in the models used in inference. In this work, we carry out a reanalysis of GW231123 using a physically complete model, accounting for both spin precession and eccentricity. Our analysis shows that this event does not exhibit strong evidence for eccentricity and the exclusion of eccentricity has minimal impact on inference. Furthermore, for GW231123-like systems, even eccentricities as large as $0.15$ at $10$ Hz do not yield a confident nonzero eccentricity measurement. Through a zero-noise injection recovery study, we show that the observed discrepancies in the parameter estimates can be explained by disagreement in the waveform models at strong spin precession, with the degree of parameter bias in the zero-noise runs being comparable to that observed for the real signal. We also show that inference performed with an eccentric, aligned-spin waveform model can yield a confident nonzero eccentricity measurement due to the degeneracy between eccentricity and spin precession. Bayesian model selection, however, rules out this interpretation in favor of the eccentric, spin precessing hypothesis, which supports zero eccentricity---a conclusion we confirm with additional zero-noise injection-recovery tests.
\end{abstract}
\maketitle
\section{Introduction}
On November 23, 2023, the LIGO observatories \cite{Aasi_2015} at Hanford and Livingston detected the gravitational-wave (GW) event GW231123 \cite{theligoscientificcollaboration2025gwtc40updatinggravitationalwavetransient,Abac_2025}, which, under the binary black hole hypothesis corresponds to the most massive system observed to date. The initial analysis by the LIGO--Virgo--KAGRA (LVK) Collaboration inferred source-frame masses of $137^{+22}_{-17} M_\odot$ and $103^{+20}_{-52} M_\odot$, placing both black hole masses within or near the theorized pair-instability mass gap (roughly $60-130 M_\odot$) \cite{Farmer_2019,Farmer_2020,Woosley_2021,hendriks2023pulsationalpairinstabilitysupernovaegravitationalwave}. In addition, the LVK analysis estimated high component spin magnitudes, with values of approximately 
$0.9$ and $0.8$ for the two black holes. The combination of large masses and high spins challenges standard stellar-evolution formation pathways, positioning GW231123 as one of the key probes of black hole formation physics. The astrophysical significance of this event has motivated numerous follow-up studies exploring alternative formation scenarios, including hierarchical mergers \cite{stegmann2025resolvingblackholefamily,li2025gw231123productsuccessivemergers,li2025hierarchicalmergerscenariogw231123,passenger2025gw231123hierarchicalmerger}, accretion driven growth in dense star clusters \cite{fulya_paper_accretion}, mergers within active galactic nucleus disks \cite{bartos2025accretionneedblackhole,delfavero2025prospectsformationgw231123agn}, primordial black holes \cite{yuan2025gw231123massgapevent,deluca2025gw231123possibleprimordialblack}, and the evolution of massive stars under specific conditions such as low metallicity \cite{gottlieb2025spinninggapdirecthorizoncollapse,tanikawa2025gw231123formationpopulationiii}, moderate magnetic fields \cite{gottlieb2025spinninggapdirecthorizoncollapse}, or high spins \cite{croon2025stellarphysicsexplaingw231123,popa2025massiverapidlyspinningbinary}. More exotic possibilities have also been considered, including cosmic strings \cite{cuceu2025gw231123binaryblackhole}, gravitational lensing \cite{Abac_2025,Goyal:2025AcrossUniverse,Shan:2025P2500747,LVK2025GWTC4Lensing,ChanEtAl2025GW231123DINGO,ChakrabortyMukherjee2025GW231123}, and even the presence of overlapping signals \cite{Hu:2025GW231123OS}.

Given the wide range of proposed formation scenarios and the event’s potential implications for black hole astrophysics, accurately determining its source properties is crucial. However, the LVK analysis revealed that these properties are subject to large systematic uncertainties. Different waveform models yield significantly different estimates of parameters such as the total mass, mass ratio, spin magnitudes, and luminosity distance. These discrepancies may stem from unmodeled eccentric features in the signal or from mismodeling of signals in the high spin or strong spin precession regime. The waveform models used in the LVK analysis assumed a quasicircular binary, whereas eccentric signatures are plausible if the system formed through a dynamical formation channel \cite{Zevin_2017,Rodriguez_2018,Gond_n_2019,PhysRevD.97.103014,Zevin_2019}. Additional evidence for missing physics comes from an independent machine learning based reconstruction \cite{chatterjee2025machinelearningconfirmsgw231123}, which was found to agree more closely with model agnostic reconstructions than with the quasicircular waveform models, suggesting that quasicircular waveform descriptions may not capture all relevant physical effects. Furthermore, a reanalysis of the GWTC-4.0 catalog \cite{Xu:2025GWTC4PE} using state-of-the-art phenomenological waveform models \cite{PhysRevD.105.084039,PhysRevD.105.084040,hamilton2025phenomxpnrimprovedgravitationalwave,planas2025timedomainphenomenologicalmultipolarwaveforms}, including an eccentric aligned-spin model, found that GW231123 shows a preference for an eccentric interpretation when compared to a quasicircular aligned-spin hypothesis; however, this preference is not supported when a quasicircular spin-precessing hypothesis is considered. In addition, the waveform models used in the LVK analysis lack calibration against numerical relativity simulations for generic spins exceeding $\sim 0.8$, a particularly relevant limitation given that the inferred median primary spin of this event lies above this threshold. Moreover, all waveform models used in the LVK analysis, except for {\tt NRSur7dq4} \cite{Varma_2019} and {\tt IMRPhenomXO4a} \cite{Thompson_2024}, are calibrated only for aligned-spin configurations.

Motivated by the possibility of unmodeled eccentric features in the signal, particularly when accounting for spin precession, and to investigate the sources of systematic errors that may be responsible for the parameter uncertainties reported under the binary black hole hypothesis, we carry out an in-depth reanalysis of this event. To this end, we employ an inspiral–merger–ringdown waveform model capable of describing eccentric and spin precessing binaries, {\tt TEOBResumS-Dalí} \cite{PhysRevD.110.024031,Chiaramello_2020,Nagar_2021,nagar2024effectiveonebodywaveformmodelnoncircularized,nagar2025effectiveonebodywaveformmodelnoncircularized}, previously used in the eccentric, spin precessing analyses of GW150914 \cite{gamba2025revisitinggw150914nonplanareccentric} and GW200105 \cite{jan2025gw200105detailedstudyeccentricity}.
Using this model, we estimate the eccentricity at 10 Hz to be $0.062^{+0.063}_{-0.062}$ and find that zero eccentricity lies within the $90\%$ highest-density interval. To investigate whether the exclusion of eccentricity in the inference could explain the discrepancies,  we also analyze the data under the quasicircular hypothesis using the same model. Furthermore, we perform zero-noise injection–recovery studies to determine the eccentricity threshold for GW231123-like systems above which we can make confident measurements of nonzero eccentricity, and unmodeled eccentricity begins to affect parameter estimation.  In additional injection–recovery tests, we investigate whether the discrepancies observed in the analysis of GW231123 can be attributed to mismodeling at high spin and strong spin precession regime. Finally, to probe a potential degeneracy between eccentricity and spin precession \cite{Ramos_Buades_2020,Romero_Shaw_2023,Xu_2023,Divyajyoti_2024,divyajyoti2025biasedparameterinferenceeccentric}  in short-duration, high-mass signals, we analyze the event under an eccentric, aligned-spin hypothesis and assess whether a confident yet biased inference of nonzero eccentricity could arise.

\section{Measuring eccentricity}
\begin{figure*}
    \centering
    \includegraphics[scale=0.41]{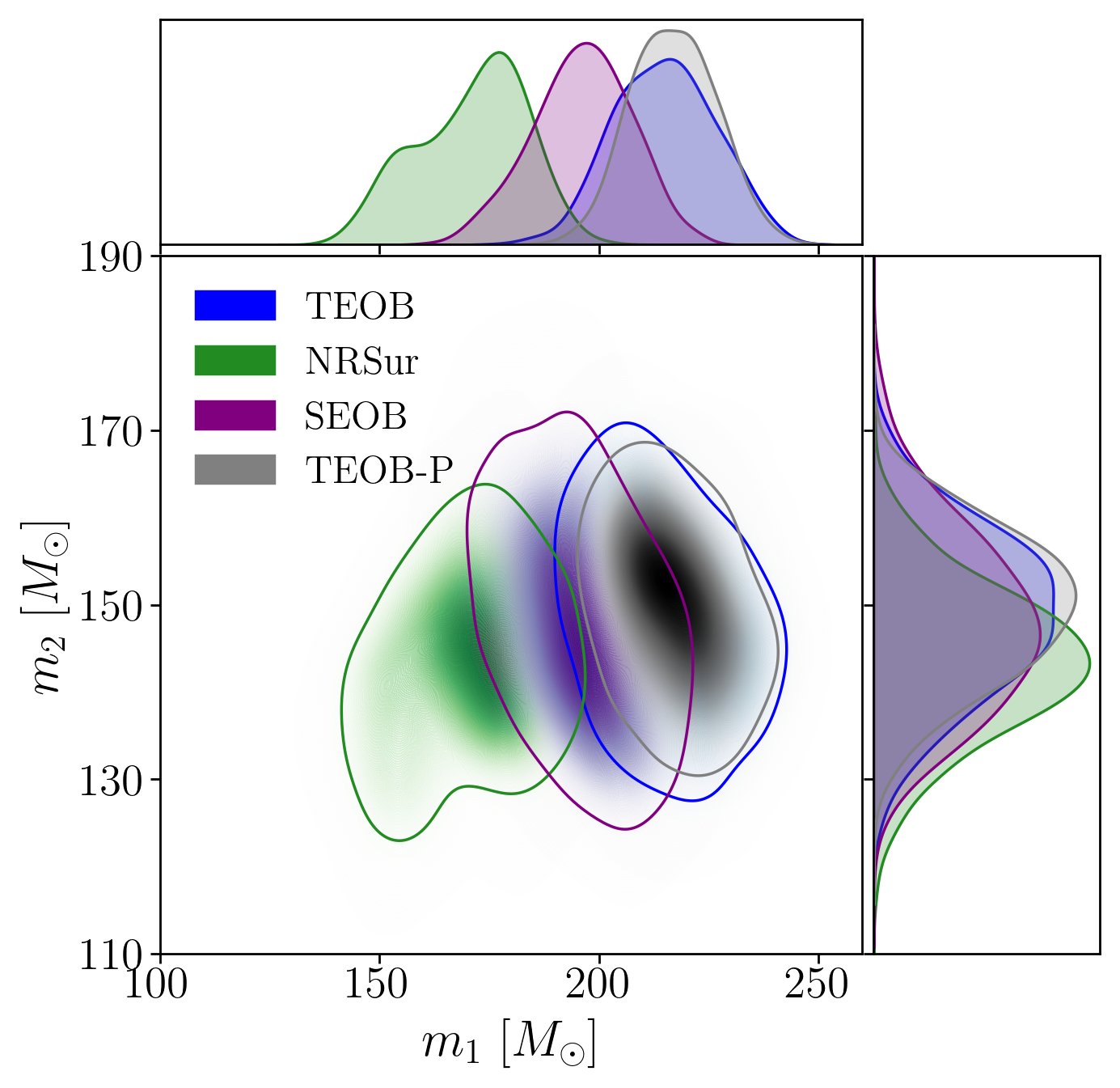}
    \includegraphics[scale=0.41]{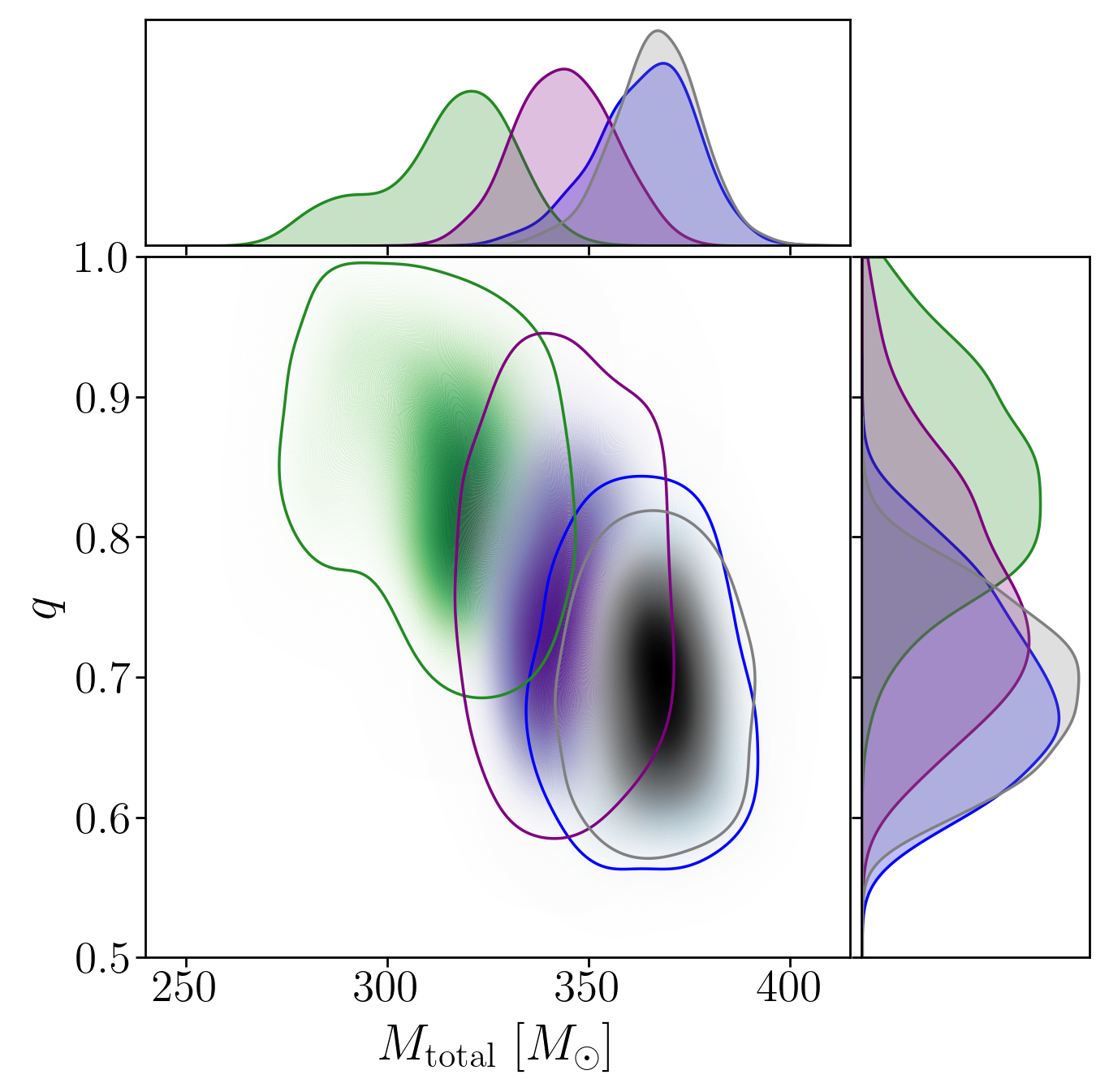}
    \includegraphics[scale=0.41]{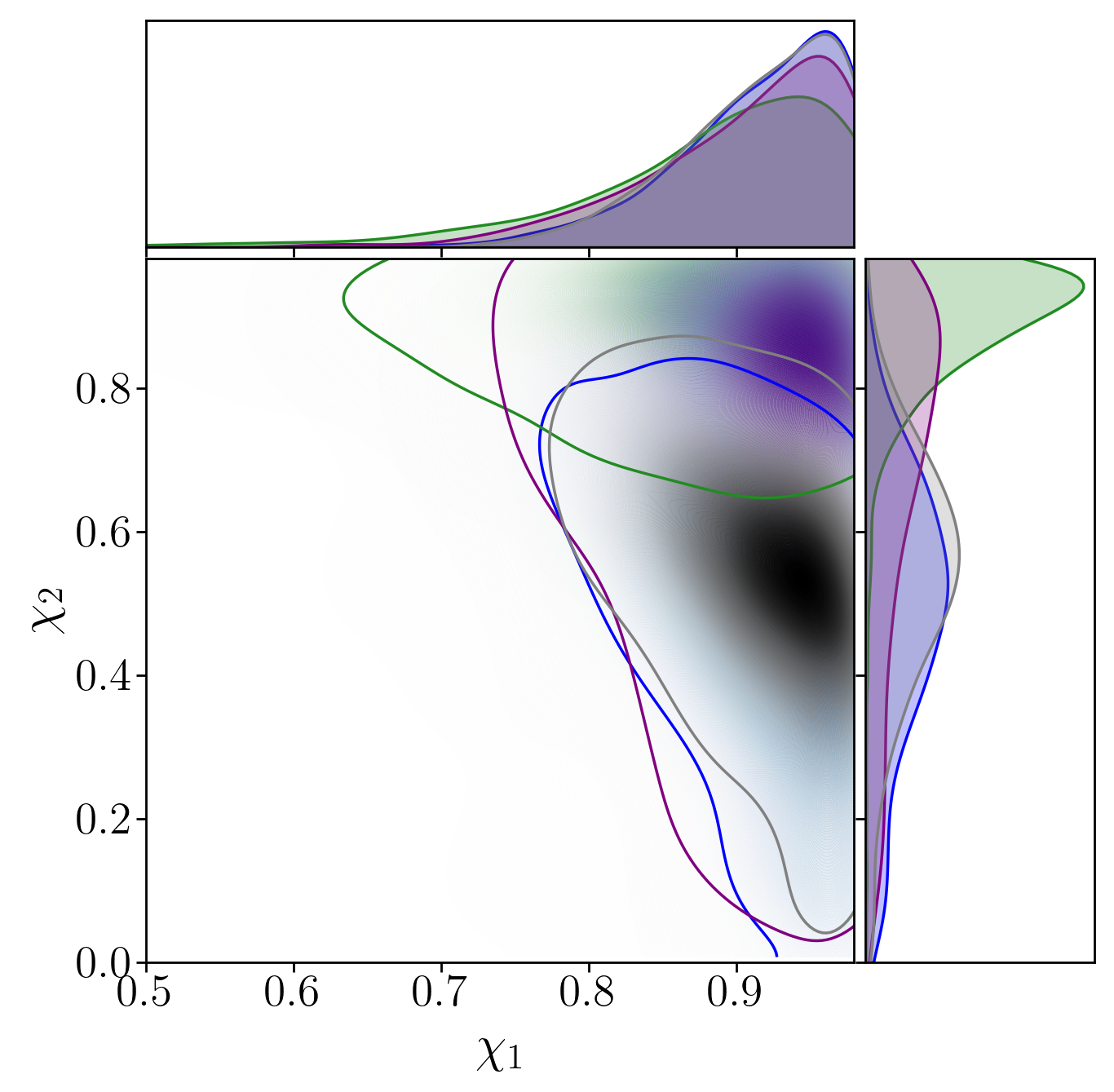}
    \caption{GW231123 results: One- and two-dimensional marginal posterior distributions for the detector-frame component masses (\textit{left}), the total mass and mass ratio (\textit{middle}), and the spin magnitudes (\textit{right}) obtained using the three waveform models (TEOB, NRSur, SEOB). Each contour shows the $90\%$ credible intervals for the joint two-dimensional marginal posterior distribution.  {\tt TEOBResumS-Dal\'{\i}} is used to analyze this event under two hypotheses: eccentric, spin precessing (TEOB) and quasicircular, spin precessing (TEOB-P).}
    \label{fig:model_comparison}
\end{figure*}
\begin{table*}
\centering
\addtolength{\tabcolsep}{3pt} 
     \begin{tabular}{c |c |c |c |c |c |c |c |c |c}
     \hline
      Model & $m_1/\text{M}_\odot$ & $m_2/\text{M}_\odot$  & $M_{\text{total}}/\text{M}_\odot$  & $q$ & $\chi_1$ & $\chi_2$ & $\chi_{\text{eff}}$ & $\chi_{\text{p}}$ & $D_L/\text{Mpc}$  \\[0.5ex] 
\hline\hline
{TEOB} & $215.0^{+19.1}_{-18.5}$ & $149.1^{+14.4}_{-15.7}$ &
$365.0^{+17.8}_{-23.3}$ & $0.69^{+0.12}_{-0.09}$ & $0.93^{+0.06}_{-0.12}$ &
$0.50^{+0.25}_{-0.40}$ & $0.43^{+0.08}_{-0.09}$ & $0.71^{+0.13}_{-0.15}$ &
$3650^{+987}_{-1223}$ \\

{NRSur} & $172.9^{+16.4}_{-23.3}$ & $143.4^{+14.0}_{-14.5}$ &
$316.9^{+19.6}_{-35.0}$ & $0.84^{+0.12}_{-0.11}$ & $0.90^{+0.08}_{-0.20}$ &
$0.91^{+0.07}_{-0.21}$ & $0.23^{+0.23}_{-0.35}$ & $0.77^{+0.17}_{-0.16}$ &
$2088^{+1586}_{-1135}$ \\

{SEOB} & $196.5^{+16.9}_{-20.1}$ & $147.6^{+18.2}_{-15.7}$ &
$344.0^{+19.8}_{-19.5}$ & $0.75^{+0.15}_{-0.11}$ & $0.92^{+0.06}_{-0.16}$ &
$0.74^{+0.23}_{-0.57}$ & $0.43^{+0.14}_{-0.15}$ & $0.74^{+0.14}_{-0.16}$ &
$3374^{+1044}_{-1220}$ \\

{TEOB\text{-}P} & $216.7^{+16.3}_{-15.0}$ & $150.0^{+12.9}_{-14.3}$ &
$367.1^{+16.7}_{-18.4}$ & $0.69^{+0.09}_{-0.09}$ & $0.93^{+0.06}_{-0.12}$ &
$0.55^{+0.24}_{-0.34}$ & $0.46^{+0.09}_{-0.09}$ & $0.71^{+0.13}_{-0.12}$ &
$3720^{+981}_{-1207}$ \\
\hline\hline
     \end{tabular}
     \caption{Summary statistics for GW231123: This table reports median values together with the 
$90\%$ equal-tailed credible intervals obtained from the analysis of GW231123 using TEOB, NRSur, SEOB, and TEOB-P. For $e_{10}$, TEOB infers a median value with a $90\%$ highest-density interval of $0.062^{+0.063}_{-0.062}$. }
    \label{tab:median_values}
\end{table*}

\begin{table*}
\centering
\begin{tabular}{l
|cc
|cc
|cc
|cc
|cc
|cc
|cc
|cc
|cc}
\hline
Model &
\multicolumn{2}{c|}{$m_1$} &
\multicolumn{2}{c|}{$m_2$} &
\multicolumn{2}{c|}{$M_{\rm tot}$} &
\multicolumn{2}{c|}{$q$} &
\multicolumn{2}{c|}{$\chi_1$} &
\multicolumn{2}{c|}{$\chi_2$} &
\multicolumn{2}{c|}{$\chi_{\rm eff}$} &
\multicolumn{2}{c|}{$\chi_\text{p}$} &
\multicolumn{2}{c}{$D_L$} \\
&
real & fake &
real & fake &
real & fake &
real & fake &
real & fake &
real & fake &
real & fake &
real & fake &
real & fake \\
\hline\hline
NRSur &
-2.80 & -2.68 &
-0.48 &  0.31 &
-2.86 & -2.69 &
 1.65 &  2.18 &
-0.24 & -0.89 &
 1.23 &  1.23 &
-2.42 &  0.25 &
 0.43 &  0.56 &
-1.49 & -1.68 \\
SEOB  &
-1.29 & -1.25 &
-0.18 & -0.10 &
-1.32 & -1.44 &
 0.66 &  0.72 &
-0.04 &  0.01 &
 0.64 &  0.76 &
-0.30 &  0.73 &
 0.26 &  0.20 &
-0.32 & -0.52 \\
\hline\hline
\end{tabular}
\caption{Normalized parameter biases for NRSur and SEOB relative to TEOB-P, for both the real GW231123 analysis (``real'') and the synthetic injection studies (``fake''). For each parameter $X$, the normalized bias is defined as $(\tilde{X}_M - \tilde{X}_{\mathrm{TEOB\text{-}P}})\big/[\tfrac{1}{2}(\Delta_+ + \Delta_-)]$, where $\tilde{X}$ denotes the posterior median. The quantities $\Delta_{\pm}$ are the upper and lower 90\% credible interval half-widths of the TEOB-P posterior.}
\label{tab:normalized_bias}
\end{table*}

We perform a Bayesian analysis of the GW strain data for GW231123 \cite{Abac_2025,GWOSC_GW231123_135430} from the LIGO Hanford and Livingston observatories using the {\tt RIFT} \cite{PhysRevD.92.023002,lange2018rapidaccurateparameterinference,wofford2023expandingriftimprovingperformance,wagner2025narrowingriftfocusedsimulationbasedinference} parameter estimation algorithm. The likelihood is evaluated using $8$s of data, and the frequencies considered in the evaluation lie in the range $20-448$ Hz, matching the settings used in the original LVK analysis. The power spectral densities used for both detectors are identical to those employed in the original LVK analysis. We do not, however, marginalize over detector calibration uncertainties since calibration curves are mostly consistent with zero \cite{theligoscientificcollaboration2025opendataligovirgo}. The prior distributions employed in the construction of posterior distributions are the same as those used in the original LVK analysis, with the addition of uniform priors on both the eccentricity and the anomaly parameter. Furthermore, both eccentricity $e_{10}$ and effective precession spin parameter $\chi_\text{p}$ \cite{Schmidt_2015} are reported at a reference frequency of $10$ Hz.

We use the {\tt TEOBResumS-Dal\'{\i}} \cite{PhysRevD.110.024031} (TEOB) waveform model to analyze the GW data, enabling us to account for spin precession and eccentricity simultaneously.  For comparison, we further analyze the data using the quasicircular, spin precessing waveform models {\tt NRSur7dq4} \cite{Varma_2019} (NRSur) and {\tt SEOBNRv5PHM} \cite{ramosbuades2023seobnrv5phmgenerationaccurateefficient} (SEOB). We find our results from these two models to be consistent with those made publicly available by the LVK Collaboration \cite{Abac_2025}. 
For each waveform model, we use all $\ell \leq 4$ GW modes provided by the waveform model and generate them starting from a $(2,2)$ mode frequency of $10$ Hz.

The results of our analysis using all three models are presented in Fig.~\ref{fig:model_comparison} and Table~\ref{tab:median_values}. In the \textit{left} panel, we find that TEOB favors a higher detector-frame primary mass $m_1$ compared to the other two models, with only a slight overlap with the SEOB posterior and minimal overlap with the NRSur posterior. All three models, however, show substantial overlap in the posteriors for detector-frame secondary mass $m_2$. These differences propagate into the total detector-frame mass $M_\text{total}$ and mass ratio $q=m_2/m_1$ posteriors shown in the \textit{middle} panel, where TEOB prefers a lower $q$ and higher $M_\text{total}$ than the other two models. In the \textit{right} panel, we find TEOB shows good agreement with SEOB and NRSur for the primary spin magnitude $\chi_1$, but it favors a lower secondary spin magnitude $\chi_2$. However, all three models produce consistent posteriors for  $\chi_\text{p}$, as shown in Table~\ref{tab:median_values}.  We note that while NRSur infers a smaller median luminosity distance $D_L$, its positively skewed posterior spans the full distance range favored by TEOB and SEOB. Furthermore, we find that the $e_{10}$ posterior, shown in Fig.~\ref{fig:eccentricity_comparison}, supports values across the full range $0.0$--$0.20$ explored in our analysis. The inferred median value and associated $90\%$ highest-density interval is $e_{10} = 0.062^{+0.063}_{-0.062}$.  Since zero eccentricity lies within this interval, we conclude that the data does not provide significant evidence for nonzero eccentricity. To assess what level of eccentricity could be confidently recovered in a GW231123-like system, we take the maximum-likelihood point from the TEOB analysis (listed in Table~\ref{tab:max_lnL_point}), inject it into a zero-noise realization, and repeat the analysis while varying the injected eccentricity across $e_{10} = \{0.0, 0.1, 0.15\}$. As shown in Fig.~\ref{fig:eccentricity_comparison}, the resulting posteriors retain substantial support for $e_{10}=0.0$ for all three injections. Only the 
$e_{10}=0.15$ injection yields a posterior with a distinct peak away from zero and a $90\%$ highest-density interval that excludes zero. Taken together, these results indicate that, at least up to $e_{10}=0.15$, a confident nonzero eccentricity measurement cannot be achieved for GW231123-like events.

We also quantify the data’s preference for each model by computing the natural logarithm of the Bayes factor, $\ln \text{BF}$, relative to the TEOB analysis. We find $\ln \text{BF}\sim0.7$ between SEOB and TEOB, in support of SEOB, and $\ln \text{BF}\sim-11$ between NRSur and TEOB, in support of TEOB.

\begin{figure}
    \centering
    \includegraphics[scale=0.5]{eccentricity_comparison.png}
    \caption{Eccentricity posterior distributions for GW231123 and GW231123-like synthetic injections (part I: measuring eccentricity):
One-dimensional marginal posterior distribution for $e_{10}$ obtained by analyzing GW231123 with TEOB. For comparison, we also show posteriors for zero-noise injections at the TEOB maximum-likelihood parameters with three different injected eccentricities. Only the injection with $e_{10}=0.15$ yields a posterior whose $90\%$ highest-density interval excludes zero and displays a distinct peak away from zero. The dashed vertical lines represent the $90\%$ high-density intervals.}
    \label{fig:eccentricity_comparison}
\end{figure}

\section{Addressing waveform systematics}
In this section, we investigate several potential sources of systematic error that may explain the discrepancies observed in the analysis of GW231123. We begin by assessing the impact of neglecting eccentricity in the inference. We then evaluate the consequences of employing waveform models at spin magnitudes that exceed their calibration ranges, as well as the impact of differences in waveform models at strong spin precession. Finally, we investigate whether analyzing the signal under the eccentric, aligned-spin hypothesis could yield a nominally confident nonzero eccentricity due to the degeneracy between eccentricity and spin precession observed in short-duration, high-mass signals. 

\textit{Exclusion of eccentricity}: If eccentric features are present in a signal, quasicircular, spin precessing waveform models can produce inconsistent parameter estimates due to model dependent interactions with the unmodeled features. In order to assess the impact of excluding eccentricity in the analysis of GW231123, we analyze the data using the {\tt TEOBResumS-Dal\'{\i}} waveform model under the quasicircular, spin precessing assumption (TEOB-P), and compare the results with those obtained from full eccentric, spin precessing TEOB analysis.  The results obtained under this assumption are presented in Fig.~\ref{fig:model_comparison}, alongside those from TEOB analysis. Across all three panels, the TEOB-P posteriors are almost identical to those from the TEOB analysis, indicating that excluding eccentricity does not significantly alter the inferred parameters. These observations are further supported by Table~\ref{tab:median_values}, which reports the inferred median values and their $90\%$ equal-tailed credible intervals. Furthermore, we observe only a negligible difference between the maximum-likelihood values obtained from the two analyses. To quantify the data’s preference regarding eccentricity in the analysis, we compute $\ln \text{BF}$ between TEOB-P and TEOB and obtain a value of $\sim 0.7$, indicating that the model under the two hypotheses fits the data almost equally well, with only a marginal preference for the quasicircular, spin precessing hypothesis. Taken together, these findings suggest that excluding eccentricity does not significantly affect the analysis of GW231123 and hence is unlikely to be the source of discrepancies observed in the parameter estimates.

\begin{figure*}
    \centering
    \includegraphics[scale=0.41]{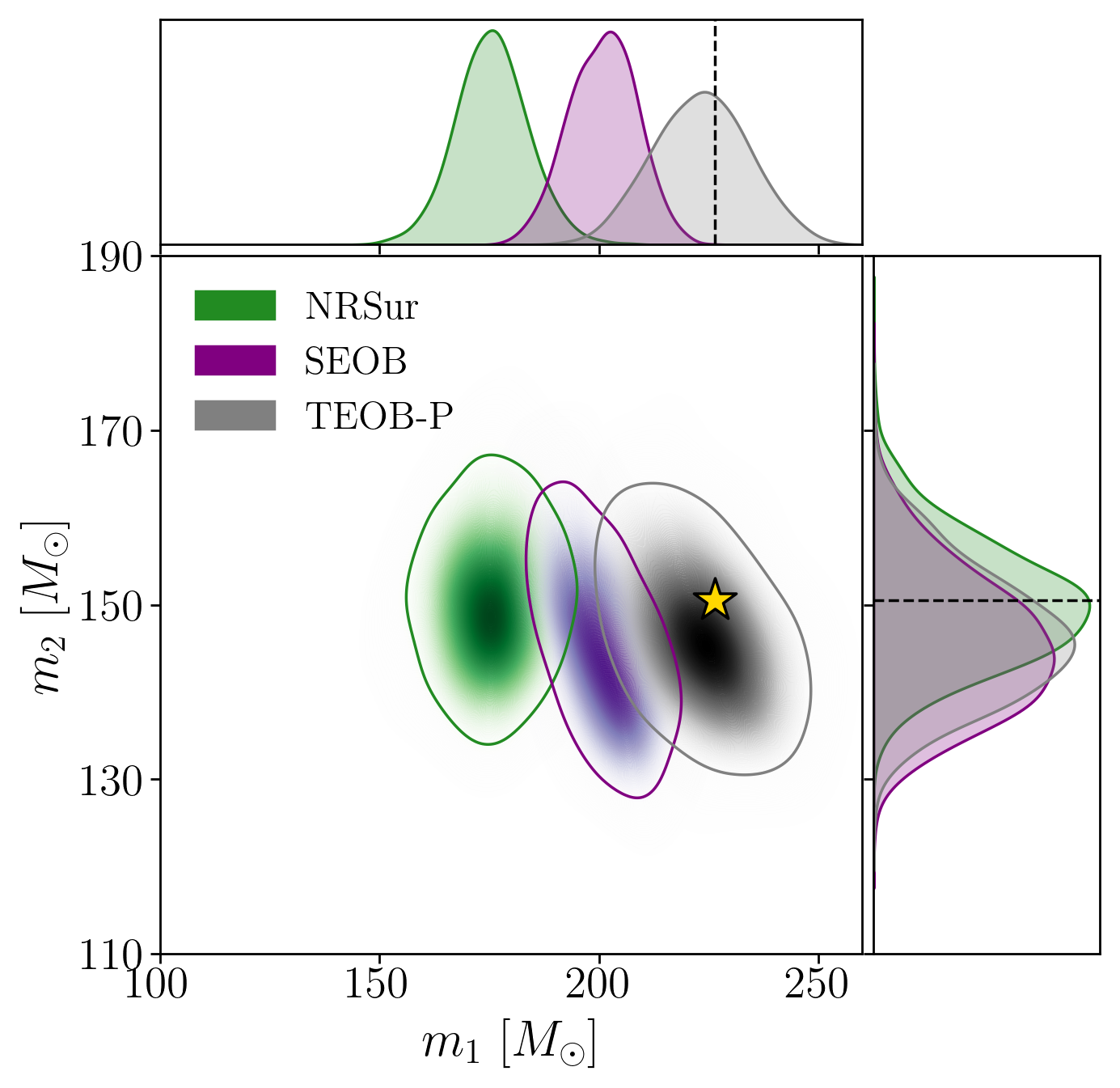}
    \includegraphics[scale=0.41]{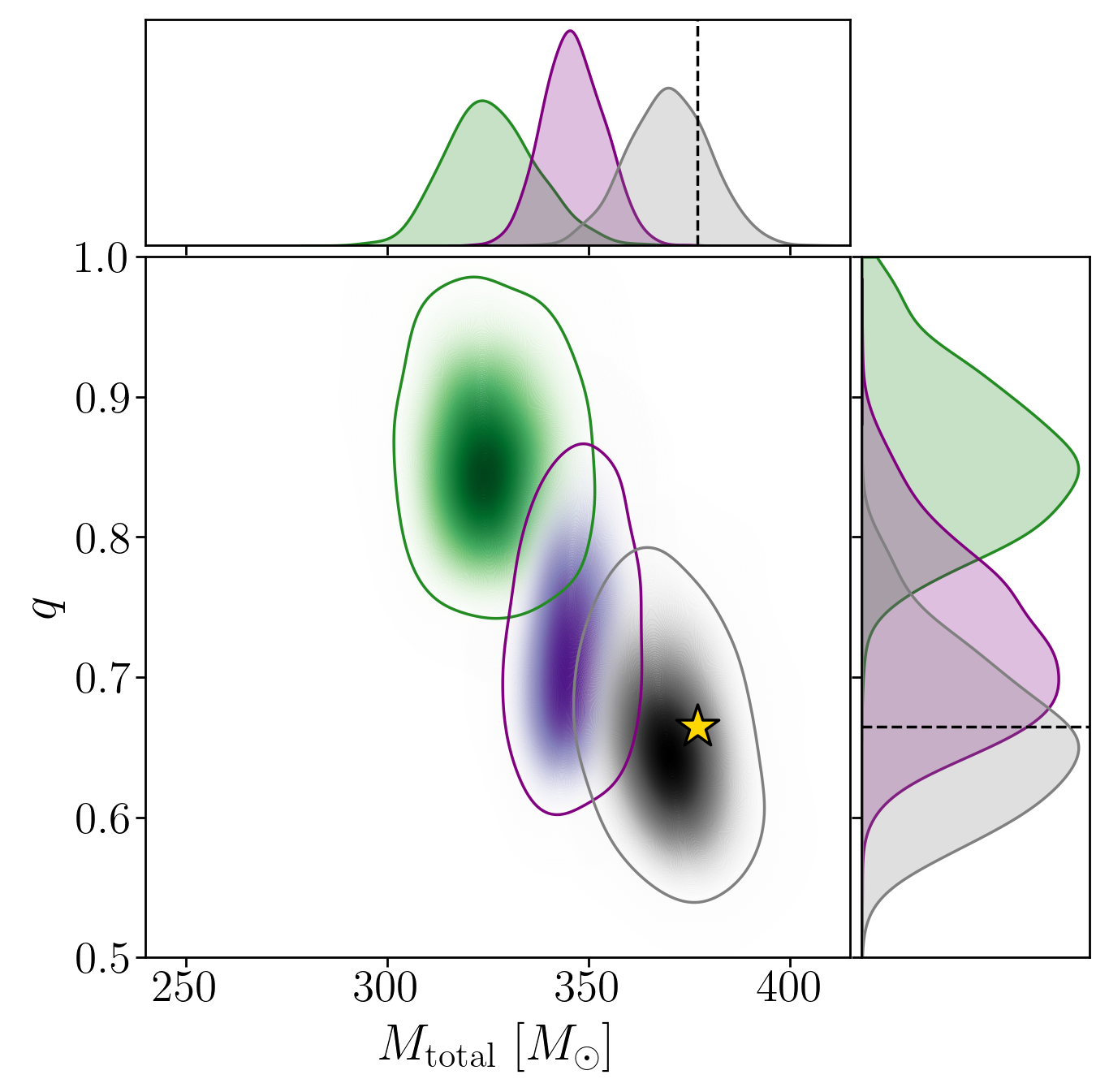}
    \includegraphics[scale=0.41]{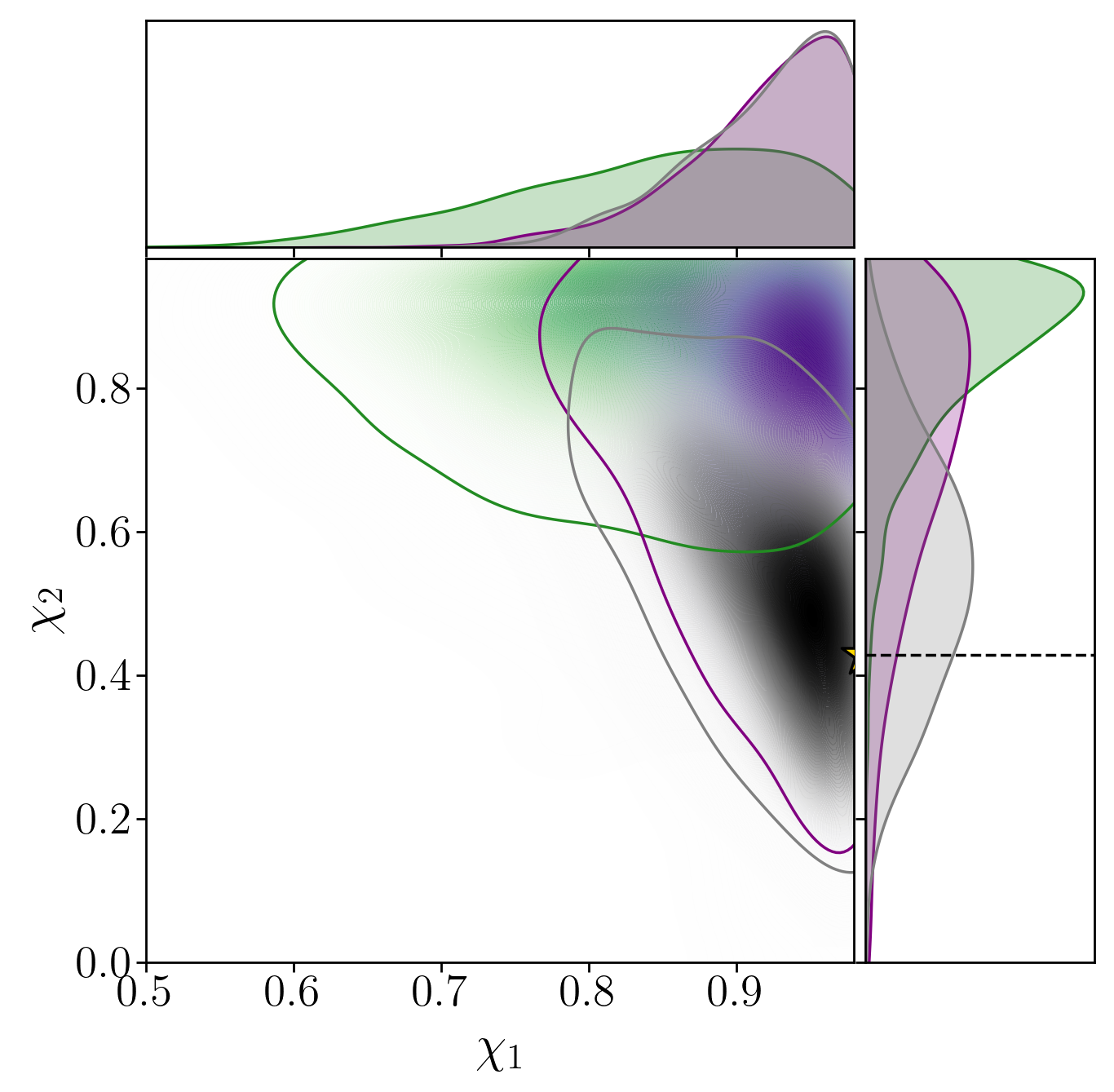}
    \caption{Impact of systematics due to strong spin precession for GW231123-like signals: One- and two-dimensional marginal posterior distributions for the detector-frame component masses (\textit{left}), the total mass and mass ratio (\textit{middle}), and the spin magnitudes (\textit{right}) obtained using the three waveform models (TEOB, NRSur, SEOB). This figure shows that the impact of waveform systematics at the strong spin precession regime is substantial and demonstrates that the waveform models disagree in the high-likelihood region relevant for GW231123. Each contour shows the $90\%$ credible intervals for the joint two-dimensional marginal posterior distribution. The dashed lines and the stars denote the injected values.}
    \label{fig:fake_injections}
\end{figure*}

\begin{figure}
    \centering
    \includegraphics[scale=0.5]{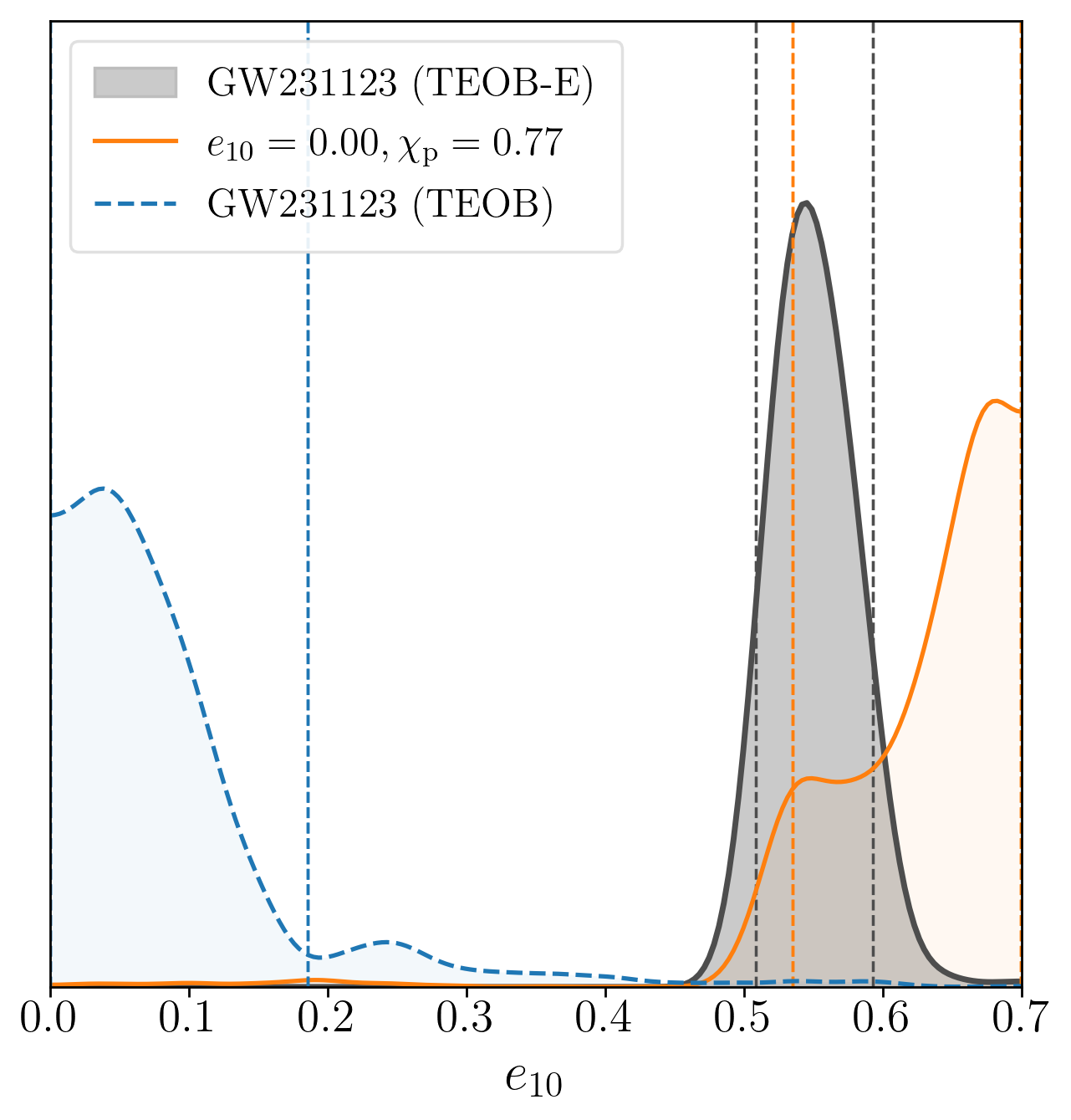}
    \caption{Eccentricity posterior distributions for GW231123 and GW231123-like synthetic injection (part II: exploring the eccentricity spin precession degeneracy):
One-dimensional marginal posterior distribution for $e_{10}$ obtained by analyzing GW231123 with TEOB-E. 
For comparison, we also show the posterior for a zero-noise injection at the TEOB maximum-likelihood point, with $e_{10}=0.0,~\chi_\text{p}=0.77$. Also shown is the posterior obtained by analyzing the event with TEOB using an extended $e_{10}$ range of $0.0-0.7$. For the zero-noise injection, the analysis with TEOB-E yields a confident, but incorrect, nonzero eccentricity measurement, demonstrating the degeneracy between eccentricity and spin precession. The dashed vertical lines indicate the $90\%$ highest-density intervals.}
    \label{fig:eccentricity_precession_degeneracy}
\end{figure}

To confirm our conclusion, we inject the maximum-likelihood point from the TEOB analysis of GW231123 at four eccentricities, $e_{10} = \{0.0, 0.05, 0.10, 0.15\}$, and analyze each injection with both TEOB and TEOB-P. For $e_{10} = 0.0$ and $0.05$, the resulting posterior distributions are almost indistinguishable. At $e_{10} = 0.10$, small differences begin to appear, though they remain insignificant, and even at $e_{10} = 0.15$ the posteriors still largely overlap. However, at $e_{10} = 0.15$, the in-plane Cartesian spin components of the two black holes exhibit significant differences, and when expressed in spherical coordinates, we find that these differences arise primarily due to differences in the inferred azimuthal angle of the in-plane spin vectors.
However, Fig.~\ref{fig:eccentricity_comparison} shows that eccentricities of this magnitude would produce a posterior with a clear peak away from zero. This gives us confidence that excluding eccentricity in the analysis of GW231123 does not meaningfully affect the inferred parameters. Interestingly, even for the $e_{10} = 0.15$ case, the $\ln \text{BF}$ is approximately $1$, favoring TEOB-P. Thus, even for relatively high eccentricity, the Bayesian evidence does not provide strong support for the eccentric, spin precessing hypothesis over a purely spin precessing one for systems like GW231123.

\textit{Operating beyond spin calibration limits}: We now assess the impact of analyzing GW231123 in a regime that lies outside the calibration limits of current waveform models. To test this, we generate a synthetic signal using TEOB-P and analyze it with TEOB-P, NRSur, and SEOB. The injection is constructed from the maximum-likelihood point from the TEOB reanalysis, with the eccentricity parameters set to zero. This point has a primary spin component magnitude above $0.8$. The purpose of this test is to determine whether the models remain mutually consistent when evaluated at these parameters: if they do, their posteriors should agree.  We first align the spin vectors with the orbital angular momentum and adjust the luminosity distance to keep the network SNR at $\sim21.8$. This allows us to test whether exceeding the calibrated spin limit, while staying within the aligned-spin modeling regime, introduces noticeable discrepancies in the recovered parameters. We find that all three waveform models yield broadly consistent posteriors. The only significant difference appears in the NRSur posterior for chirp mass, which shows a modest deviation from the TEOB-P and SEOB results, though with substantial overlap. Overall, these results suggest that simply stepping outside the calibrated spin range is unlikely to be the primary driver of the discrepancies seen in the GW231123 analysis.

To investigate whether strong spin precession—combined with primary spin magnitude outside the calibrated range—contributes to the observed discrepancies, we perform a second injection–recovery test. This time, we inject the maximum-likelihood point from the TEOB reanalysis without modifying the spin orientations, thereby retaining the original level of spin precession, and reanalyze the resulting synthetic signal with TEOB-P, NRSur, and SEOB. At these parameters, the system is exhibiting strong spin precession, with $\chi_\text{p} = 0.77$. As shown in Fig.~\ref{fig:fake_injections}, waveform systematics become substantially more pronounced. Comparing Fig.~\ref{fig:model_comparison} with Fig.~\ref{fig:fake_injections}, we find that the extent and pattern of the resulting biases resemble those seen in the real-data analysis of GW231123 for most of the parameters, a correspondence further supported by normalized bias values in Table~\ref{tab:normalized_bias}, which exceed one for multiple parameters. Both NRSur and SEOB favor a lower $m_1$ than TEOB, while all three models yield consistent posteriors for $m_2$. As in the real-data analysis, NRSur prefers mass ratios closer to unity and a lower $M_\text{total}$, and both NRSur and SEOB infer higher values of $\chi_2$ despite the true value being $\sim 0.4$. However, the bias behavior observed in the real-data analysis for the Cartesian spin components is not reproduced in the injection study. Taken together, these similarities do not, on their own, prove that strong spin precession is solely responsible for the discrepancies, but they strongly suggest that spin precession in this region of parameter space plays a significant role in the systematic differences observed across waveform models.

\textit{Eccentricity spin-precession degeneracy}: Both eccentricity and spin precession produce similar signatures in waveforms, as they each induce GW amplitude and phase modulations \cite{PhysRevD.74.122001,Schmidt_2015,PhysRevD.97.024031}. This similarity can become problematic for heavy systems such as GW231123, where only a small number of GW cycles are observed. In general, distinguishing these two physical effects requires that the observed signal spans a significant fraction of the characteristic timescale \cite{Morras_2025} associated with each effect. When the signal is too short to resolve these timescales, the effects can become degenerate \cite{PhysRevLett.126.201101}. Consequently, for such short signals, an eccentric, aligned-spin waveform model can incorrectly attribute spin precession induced modulations to eccentricity \cite{snehal_degeneracy}. To test this possibility, we analyze GW231123 using {\tt TEOBResumS-Dal\'{\i}} under the eccentric, aligned-spin hypothesis (TEOB-E) and examine whether it yields a confident nonzero eccentricity measurement. Because the inferred $e_{10}$ posterior accumulated near the original upper bound of $e_{10}=0.2$, we increase the upper limit to $0.7$. After doing so, we find the resulting $e_{10}$ posterior is well constrained, as shown in Fig.~\ref{fig:eccentricity_precession_degeneracy}, with a median and the corresponding $90\%$ highest-density interval of $0.548^{+0.045}_{-0.040}$. Moreover, zero eccentricity is completely excluded from the posterior. 

To directly compare these results with those obtained using TEOB and to assess whether this high eccentricity region is supported under the eccentric, spin precessing hypothesis, we reanalyze the event with TEOB using the expanded $e_{10}$ range of $0.0-0.7$. The results, also shown in Fig.~\ref{fig:eccentricity_precession_degeneracy}, confirm our earlier findings under the narrower prior: Zero eccentricity remains included in the posterior, and we infer a median value of  $0.066^{+0.120}_{-0.066}$. Furthermore, the maximum log-likelihood under the eccentric, aligned-spin hypothesis is smaller by approximately $15$, and the corresponding $\ln \text{BF} \sim -26$ between TEOB-E and TEOB strongly supports the eccentric, spin precessing hypothesis. This strong preference for the eccentric, spin precessing hypothesis, together with the fact that the $e_{10}$ posterior shows strong support for zero, indicates that GW231123 does not provide evidence for eccentricity. Instead, the apparently confident nonzero eccentricity inferred under the eccentric, aligned-spin hypothesis arises from the degeneracy between eccentricity and spin precession in such short signals. To confirm this interpretation, we analyze the zero-noise maximum-likelihood TEOB injection, which has $e_{10}=0.0$ and $\chi_\text{p} = 0.77$, using TEOB-E. As shown in Fig.~\ref{fig:eccentricity_precession_degeneracy}, TEOB-E erroneously finds strong support for nonzero eccentricity, yielding an inferred value of $0.642^{+0.057}_{-0.107}$ and being ruled out with $\ln \text{BF} \sim -21$.

\section{Conclusions}
In this work, we reanalyzed GW231123 with two primary goals: measuring its eccentricity and investigating potential sources of systematic error that may explain the discrepancies observed in the LVK Collaboration's analysis of this event \cite{Abac_2025}. Using the physically complete TEOB waveform model, which can generate waveforms for binary black holes in generic (eccentric and spin precessing) orbits, we found that this event does not exhibit strong evidence for eccentricity. We inferred a median value of
$e_{10} = 0.062^{+0.063}_{-0.062}$, with zero eccentricity included in the $90\%$ highest-density interval and the posterior distribution retaining significant support for zero eccentricity. We then compared the TEOB results with our results obtained using the two quasicircular, spin precessing models employed in the LVK Collaboration’s analysis of this event, namely NRSur and SEOB. We found that all three waveform models yield different parameter estimates, especially for parameters such as total mass, mass ratio, secondary spin magnitude, and luminosity distance. However, their posterior distributions show non-negligible degree of overlap. Relative to TEOB, we obtained the log Bayes factors ($\ln \text{BF}$) of $0.7$ for SEOB and $-11$ for NRSur.

To investigate the origin of the differences in parameter estimates, we first analyzed the data for GW231123 under the quasicircular, spin precessing hypothesis using TEOB-P to assess whether neglecting eccentricity is the source of the discrepancies. We found that the inferred posterior distributions show no significant difference compared to those obtained with TEOB under the eccentric, spin precessing hypothesis. Furthermore, Bayesian model selection could not distinguish between the eccentric, spin precessing and quasicircular, spin precessing hypotheses, with the $\ln \text{BF}$ between TEOB-P and TEOB being $0.7$. This demonstrates that excluding eccentricity does not significantly affect the inferred parameters for this event and is hence unlikely to be the source of discrepancies for this event. We confirmed this conclusion through zero-noise injection–recovery studies at the maximum-likelihood point from the TEOB analysis of GW231123. These studies showed that for eccentricities up to $0.1$, the posterior distributions obtained using the two hypotheses show no significant difference. Furthermore, the injected $e_{10}$ must exceed $0.1$ for zero eccentricity to lie outside the $90\%$ highest-density interval.  Moreover, for all injected eccentricities explored in this study, up to $e_{10}=0.15$, the inferred posterior distributions continued to retain substantial support for zero eccentricity.

We then performed a zero-noise injection–recovery study at the maximum-likelihood TEOB point, but with 
eccentricity parameters set to zero. This point is characterized by a highly spinning primary black hole and strong spin precession, with the primary spin magnitude exceeding the current model calibration limits. Our analysis of this injection using NRSur and SEOB showed significant biases in parameter estimates, with the posterior shifts relative to TEOB-P comparable to those observed in the analysis of real data. This indicates that the waveform models behave inconsistently in the high-likelihood region of parameter space favored by this event and that such inconsistencies are, at minimum, contributing to the discrepant results obtained for this event. In contrast, analyzing a zero-noise injection with the same parameters but with both spins aligned with the orbital angular momentum showed that exceeding the calibrated spin range of current waveform models in the aligned-spin modeling limit does not produce significant discrepancies. In other words, while cross-model systematics for nonprecessing binaries are modest even for the large spins considered here, allowing for both high spin and spin precession leads to much more substantial cross-model systematics.

We also demonstrated that analyzing this event with an eccentric, aligned-spin waveform model can yield a nominally confident, nonzero eccentricity measurement. However, Bayesian model selection strongly favors the eccentric, spin precessing hypothesis, under which the posterior supports zero eccentricity. This indicates that the apparent nonzero eccentricity arises from a degeneracy between eccentricity and spin precession. We confirmed this conclusion through an eccentric, aligned-spin recovery of a zero-noise quasicircular spin precessing injection at the maximum-likelihood point from the TEOB reanalysis of the event, which again led to a confidently nonzero eccentricity estimate---despite the injected signal being quasicircular and strongly precessing.

The analysis of GW231123 highlights several challenges that must be addressed to extract the full scientific potential of such signals. A central issue is that current waveform models disagree when the system exhibits strong spin precession, as is the case for GW231123, whose high-likelihood region corresponds to highly precessing configurations. These disagreements can introduce significant biases in the inferred parameters, even at a moderate signal-to-noise ratio of $21.8$. Addressing these shortcomings will require improvements in how waveform models treat spin precession. Another key challenge is the development of waveform models capable of capturing the coupled dynamics of spin precession and eccentricity in a self-consistent manner, so that limitations in modeling one physical effect are not misinterpreted as evidence for—or against—a physical effect. A further limitation is that current inspiral-merger-ringdown eccentric waveform models typically include eccentricity only in the inspiral and assume circularization in the merger and ringdown. For short-duration signals like GW231123, this can be consequential. Overcoming these issues will require high-accuracy numerical relativity waveforms with broad coverage of highly spin precessing, eccentric, and eccentric, spin precessing binaries to support the calibration and validation of future models. As detector sensitivity improves and GW observations become more frequent, events with similarly challenging properties will become increasingly common. Closing these modeling gaps is therefore essential for achieving unbiased parameter inference and fully exploiting the rich astrophysical information carried by future detections.

\begin{acknowledgements}
We thank Isobel Romero-Shaw for a thorough review of the manuscript. We are grateful to Alessandro Nagar for answering questions regarding the waveform model {\tt TEOBResumS-Dalí}. We also acknowledge helpful comments from Mark Hannam, Eleanor Hamilton, Jose Ezquiaga, Juan Calderón Bustillo, Yumeng Xu, and Geraint Pratten. The posterior samples generated in our analysis, along with the configuration file required to reproduce our analysis, are available at \href{https://github.com/AasimZJan/GW231123_samples}{https://github.com/AasimZJan/GW231123$\_$samples}. A.J. and D.S. acknowledge support from NASA Grant No. 80NSSC24K0437 and NSF Grant No. PHY-2207780. 
R.O.S.  gratefully acknowledges support from NSF Award No. NSF PHY-1912632, No. PHY-2012057, No. PHY-2309172, No. AST-2206321, and the Simons Foundation. 
This material is based upon work supported by NSF's LIGO Laboratory, which is a major facility fully funded by the NSF. The authors are grateful for computational resources provided by the LIGO Laboratory and supported by NSF Grants PHY-0757058 and PHY-0823459.  
The work was done by members of the Weinberg Institute and has an identifier of UT-WI-41-2025.

\end{acknowledgements}
\section*{DATA AVAILABILITY}
The data that support the findings of this article are openly available from the Gravitational Wave Open Science Center at  \cite{GWOSC_GW231123_135430}.
\appendix
\section{MAXIMUM LIKELIHOOD PARAMETERS}
\begin{table}[h!]
\centering
\begin{tabular}{l c}
\hline
Parameter & Value \\
\hline\hline
$m_1$ [$M_\odot$] & 226.5 \\
$m_2$ [$M_\odot$] & 150.5 \\
$\chi_{1x}$ & 0.36 \\
$\chi_{1y}$ & 0.68 \\
$\chi_{1z}$ & 0.62 \\
$\chi_{2x}$ & 0.26 \\
$\chi_{2y}$ & 0.23 \\
$\chi_{2z}$ & 0.25 \\
$\chi_1$ & 0.98 \\
$\chi_2$ & 0.43 \\
$\chi_{\mathrm{eff}}$ & 0.47 \\
$\chi_{\mathrm{p}}$ & 0.77 \\
$e_{10}$ & 0.10 \\
mean anomaly [rad] & 2.83 \\
right ascension [rad] & 3.02 \\
declination [rad] & -0.02 \\
coalescence phase [rad] & 2.77 \\
inclination [rad] & 2.33 \\
polarization [rad] & 2.98 \\
$D_L$ [Mpc] & 3635 \\
coalescence time [s] & 1384782888.61 \\
\hline\hline
\end{tabular}
\caption{Maximum-likelihood parameter values obtained from analyzing GW231123 with the TEOB waveform model.}
\label{tab:max_lnL_point}
\end{table}

\bibliography{references}

\end{document}